\documentclass{elsart}
\usepackage{natbib}
\usepackage{graphicx}
\begin{document}
\runauthor{Y.~Aritomo and M.~Ohta}
\begin{frontmatter}
\title{Analysis of fusion-fission dynamics by pre-scission neutron emission
 in $^{58}$Ni+$^{208}$Pb}
\author[Dubna]{Y.~Aritomo}
\author[Kobe]{M.~Ohta}
\author[ulb]{T.~Materna}
\author[ulb]{F.~Hanappe}
\author[ires]{O.~Dorvaux}
\author[ires]{L.~Stuttge}

\address[Dubna] {Flerov Laboratory of Nuclear
Reactions, JINR, Dubna, Russia}
\address[Kobe]{Department of Physics, Konan University, 8-9-1
  Okamoto, Kobe, Japan}
\address[ulb]{Universite Libre de Bruxelles, 1050 Bruxelles, Belgium}
\address[ires]{Institut de Recherches Subatomiques, F-67037 Strasbourg Cedex,
France }

\begin{abstract}
We analyzed the experimental data of the pre-scission neutron
multiplicity in connection with fission fragments in the reaction
$^{58}$Ni+$^{208}$Pb at the incident energy corresponding to the
excitation energy of compound nucleus $E^{*}$=185.9 MeV, which was
performed by D\'{e}MoN group. The relation between the
pre-scission neutron multiplicity and each reaction process having
different reaction time is investigated. In order to study the
fusion-fission process accompanied by neutron emission, the
fluctuation-dissipation model combined with a statistical model is
employed. It is found that the fusion-fission process and the
quasi-fission process are clearly distinguished in correlation
with the pre-scission neutron multiplicity.
\end{abstract}
\begin{keyword}
superheavy elements, fluctuation-dissipation dynamics,
fusion-fission process, quasi-fission process, pre-scission
neutron multiplicity
\end{keyword}
\end{frontmatter}




\section{Introduction}

Many experiments on the induced fission at near and below Coulomb
barrier in superheavy-mass region have been done and the mass and
kinetic energy distributions of fission fragments were measured
\cite{itki01}. In these phenomena, the reaction process is
classified into several characteristic processes, they are the
fusion-fission process (FF), the quasi-fission process (QF) and
the deep inelastic collision process (DIC). Depending on the shell
effect of the compound nucleus or the composite nucleus, the deep
quasi-fission process (DQF) can be seen in a certain reaction
system \cite{ari04}. In the DQF process, mass symmetric fission
fragments are observed but no compound nucleus is formed.

Each process has its own characteristic reaction time from the
contact of the colliding partner to the scission point. This was
shown by the dynamical calculation for the time development of the
nuclear shape in terms of the Langevin equation \cite{ari04}. The
different reaction time means that each reaction process
associates the different pre-scission neutron multiplicity.

We undertake to extend our model discussed in reference
\cite{ari04}, by taking into account the effect of neutron
emission. We combine the Langevin calculation with the statistical
model which calculates the neutron emission. We apply our model to
the recent experiment, in which the pre-scission neutron
multiplicity correlated with the mass distribution of fission
fragments has been measured in the reaction $^{58}$Ni+$^{208}$Pb
at the incident energy corresponding to the excitation energy of
compound nucleus $E^{*}= 185.9$ MeV. This experiment was done by
D\'{e}MoN group \cite{dona99}.
The feature of distribution of pre-scission
neutron multiplicity $\nu_{n}$ has\ two distinct components, one
located around $\nu_{n} =4 $ and the other  around $\nu_{n} =8 $.
This structure may be sign of two mechanisms having two different
life-time of the composite system, they are, the QF process and
the FF process. By our model calculation, it can be clearly shown
the two components of the neutron multiplicity are corresponding
to the above two processes.

In this paper, we show theoretically the usefulness of the methods
of classifying the dynamical process on the basis of the
pre-scission neutron multiplicity correlated with the mass
distribution of fission fragments. This approach is a powerful
tool for investigating fusion-fission mechanism and for estimating
precise cross section relevant to the problem on superheavy
elements synthesis.

In section~2, we briefly explain our framework for the study and
the model. We discuss the pre-scission neutron multiplicity in
correlation with fission fragments in section~3. We also present
how the pre-scission neutron multiplicity depends on the strength
of the friction, the level density parameter and the neutron
binding energy used in the statistical model. In section 4, we
present a summary and further discussion to clarify the reaction
mechanism in superheavy mass region.

\section{Model}

Using the same procedure as described in reference \cite{ari04} to
investigate the fusion-fission process dynamically, we use the
fluctuation-dissipation model and employ the Langevin equation. We
adopt the three-dimensional nuclear deformation space given by
two-center parameterization \cite{maru72,sato78}. The neck
parameter $\epsilon$ is fixed to be 1.0 in the present
calculation, so as to retain the contact-like configuration more
realistically for two-nucleus collision. The three collective
parameters involved in the Langevin equation are as follows:
$z_{0}$ (distance between two potential centers), $\delta$
(deformation of fragments) and $\alpha$ (mass asymmetry of the
colliding nuclei); $\alpha=(A_{1}-A_{2})/(A_{1}+A_{2})$, where
$A_{1}$ and $A_{2}$ denote the mass numbers of the target and the
projectile, respectively.

The multidimensional Langevin equation is given as
\begin{eqnarray}
\frac{dq_{i}}{dt}&=&\left(m^{-1}\right)_{ij}p_{j},\nonumber\\
\frac{dp_{i}}{dt}&=&-\frac{\partial V}{dq_{i}}
                 -\frac{1}{2}\frac{\partial}{\partial q_{i}}
                   \left(m^{-1}\right)_{jk}p_{j}p_{k}
                  -\gamma_{ij}\left(m^{-1}\right)_{jk}p_{k}
                  +g_{ij}R_{j}(t),
\end{eqnarray}
where a summation over repeated indices is assumed. $q_{i}$
denotes the deformation coordinate specified by $z_{0}$, $\delta$
and $\alpha$. $p_{i}$ is the conjugate momentum of $q_{i}$. $V$ is
the potential energy, and $m_{ij}$ and $\gamma_{ij}$ are the
shape-dependent collective inertia parameter and dissipation
tensor, respectively. A hydrodynamical inertia tensor is adopted
in the Werner-Wheeler approximation for the velocity field, and
the wall-and-window one-body dissipation is adopted for the
dissipation tensor \cite{bloc78,nix84,feld87}. The normalized
random force $R_{i}(t)$ is assumed to be  white noise, {\it i.e.},
$\langle R_{i}(t) \rangle$=0 and $\langle R_{i}(t_{1})R_{j}(t_{2})
\rangle = 2 \delta_{ij}\delta(t_{1}-t_{2})$. The strength of
random force $g_{ij}$ is given by $\gamma_{ij}T=\sum_{k}
g_{ij}g_{jk}$, where $T$ is the temperature of the compound
nucleus calculated from the intrinsic energy of the composite
system as $E_{int}=aT^2$, with $a$ denoting the level density
parameter. The temperature-dependent potential energy is defined
as

\begin{equation}
V(q,l,T)=V_{DM}(q)+\frac{\hbar^{2}l(l+1)}{2I(q)}+V_{shell}(q)\Phi
(T),
 \label{vt1}
\end{equation}
\begin{equation}
V_{DM}(q)=E_{S}(q)+E_{C}(q),
\end{equation}
where $I(q)$ is the moment of inertia of a rigid body at
deformation $q$,  $V_{shell}$ is the shell correction energy at
$T=0$, and $V_{DM}$ is the potential energy of the finite-range
liquid drop model. $E_{S}$ and $E_{C}$ denote a generalized
surface energy \cite{krap79} and Coulomb energy, respectively. The
centrifugal energy arising from the angular momentum $l$ of the
rigid body is also considered. The temperature-dependent factor
$\Phi$ is parameterized as $\Phi=$exp$\{-aT^{2}/E_{d}\}$ following
the work of Ignatyuk et al. \cite{ign75}. The shell dumping energy
$E_{d}$ is chosen to be 20 MeV. The intrinsic energy of the
composite system $E_{int}$ is calculated for each trajectory as

\begin{equation}
E_{int}=E^{*}-\frac{1}{2}\left(m^{-1}\right)_{ij}p_{i}p_{j}-V(q,l,T),
\end{equation}

where $E^{*}$ denotes the excitation energy of the compound
nucleus, and is given by $E^{*}=E_{cm}-Q$ with $Q$ and $E_{cm}$
denoting the $Q-$value of the reaction and the incident energy in
the center-of-mass frame, respectively.

%


We take into account neutron emission in the Langevin calculation
during the fusion-fission process. The present work is the first
theoretical attempt to investigate neutron evaporation in the
dynamical calculation of the fusion-fission process in the
superheavy-mass region. Here, we assume that no neutron is emitted
in the approaching process. Using the procedure outlined by
Fr\"{o}brich et al. \cite{frob93}, in which the Langevin
calculation and the statistical model were combined, the emission
of neutrons has been coupled to the three-dimensional Langevin
equation for the fusion-fission process. Due to the strong
friction in this model, we assume that the kinetic energy of
relative motion quickly dissipates into the intrinsic energy after
contact. The thermal equilibrium in the composite system in this
calculation is realized immediately within the time order of
$10^{-22}$ s \cite{ari98}. We also assume that the particle
emissions in the composite system are limited to neutron
evaporation in the neutron-rich heavy nuclei.

The width of neutron emission is expressed as

\begin{equation}
\Gamma_n=g_{n}\frac{1}{2\pi\rho_{A}(E_{int})}
\int^{E_{int}-B_n}_{0} \sum_{l}(2l+1)T_{l}(\epsilon)
\rho_{A-1}(E_{int}-B_n-\epsilon) d\epsilon,
\end{equation}

where $g_{n}$ is the spin multiplicity of the neutron. $B_n$ and
$\epsilon$ denote the neutron separation energy and the kinetic
energy of the emitted neutron, respectively. The transmission
coefficient is presented by $T_{l}$. The nuclear level density is
denoted by $\rho(E)$ with suffixes $A$ and $A-1$ for the compound
nucleus and the residual one, respectively. $R$ is the radius of
the compound nucleus. Using the sharp cut off treatment for the
angular momentum distribution of $T_{l}$, we express Eq.~(5) as
follow:

\begin{equation}
\Gamma_n=g_{n}\frac{1}{2\pi\rho_{A}(E_{int})}\frac{2\mu
R^{2}}{\hbar^{2}}\int^{E_{int}-B_n}_{0}
\epsilon\rho_{A-1}(E_{int}-B_n-\epsilon) d\epsilon.
\end{equation}

where $\mu$ is the neutron mass \cite{vand73,ohta00}.  In the case
of the excitation energy corresponding to a high incident energy
compared with the Bass barrier height \cite{bass741}, as in the
present case, the integration can be performed approximately, if
we assume a simple form of $\rho(E)\propto e^{2\sqrt{a_{n}E}}$ for
the level density \cite{vand73}, as follows:

\begin{equation}
\Gamma_n \simeq g_{n}\frac{\mu R^{2}}{\pi \hbar^{2}
a_{n}}(E_{int}-B_{n}) \exp\{-\sqrt{\frac{a_{n}}{
E_{int}}}B_{n}\},\label{eq:31}
\end{equation}

where $a_{n}$ is the level density parameter.



The Langevin calculation is solved by using constant time steps
$\Delta t$, and during each time step, the neutron emission is
calculated by the following Monte Carlo procedure. We obtain the
lifetime of neutron emission $\tau_n=\hbar / \Gamma_n$ from
Eq.(7). The probability of neutron emission during the time
interval $\Delta t$ is described by $\frac{\Delta t}{\tau_{n}}$.
If the ratio $\frac{\Delta t}{\tau_{n}}$ (in the actual
calculation, $\Delta t < \tau_{n}$) is larger than a random number
$\xi$  $(0 \leq \xi \leq 1)$, the emission of a neutron is
registered. The kinetic energy of the emitted neutron $\epsilon
^{(i)}$ is estimated stochastically by assuming the neutron
spectrum from the composite system with the temperature $T$ to be
$\epsilon^{1/2} \exp{(-\epsilon /T )} $.  The upper index ($i$)
indicates the $i$-th neutron emission stage. Therefore, the energy
loss in the neutron evaporation process is estimated as
$-\epsilon^{(i)}-B_{n}^{(i)}$. Then we set a new intrinsic energy
$E^{(i)}_{int}=E_{int}^{(i-1)}-\epsilon^{(i)}-B_{n}^{(i)}$, where
for $i=1$, $E^{(0)}_{int}$ means the initial value of $E_{int}$.
The potential energy surface also changes as described by Eq.(2)
as the temperature decreases.
After the initial energy is reset, we proceed to the next time
step of the Langevin calculation.

\section{Results and discussion}

Using our model, we analyze the experimental data in the reaction
$^{58}$Ni+$^{208}$Pb at $E^{*}$=185.9 MeV \cite{dona99}. Figure~1
shows the experimental result, which is the distribution of the
pre-scission neutron multiplicity $\nu_{n}$ in correlation with
fission fragments whose mass number is greater than
$\frac{A}{2}-30$ and less than $\frac{A}{2}+30$.

An interesting feature of the distribution of pre-scission neutron
multiplicity is its shape having two components, which are located
around $\nu_{n}=4$ and $\nu_{n}=8$. This structure may be the sign
of a simultaneous coexistence of two mechanisms corresponding to
different life time of the composite system but to the phenomena
giving nearly the same mass fragment in the fission process. The
first one, defined as the QF process, would lead to the emission
of about 4 neutrons only and for the second one, associated here
with the FF process via a compound nucleus, would be found around
8 neutron emission. In this case, the fission process would tail
long enough time to allow the emission of nearly 8 neutrons. By
our model calculation, we try to confirm that the different
dynamical processes, i.e., the FF process and the QF process, are
giving different pre-scission neutron multiplicities in spite of
the phenomenon associated with the similar mass fragment.

\begin{figure}[h]
\centerline{
  \includegraphics[height=0.37\textheight]{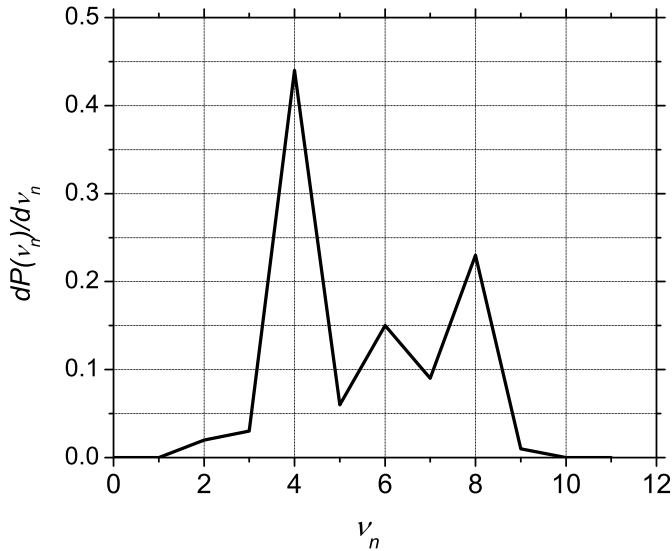}}
  \caption{Experimental pre-scission neutron multiplicity associated with fission
  fragment measurements in the reaction $^{58}$Ni+$^{208}$Pb
  at $E^{*}$=185.9 MeV \cite{dona99}.}
\end{figure}




As mentioned in the previous section, in
the calculation of neutron emission, there exist two important
parameters, the level density parameter $a_{n}$ and the neutron
binding energy $B_{n}$. Both of these parameters depend on the
deformation of the nucleus. In the dynamical calculation itself,
another unknown factor is the dissipation energy $E_{diss}$, which
is converted from the relative kinetic energy of the colliding
system to the intrinsic energy during the approaching process. In
this section, first we present the structure appeared in our model
calculation of the pre-scission neutron multiplicity. Secondly, we
discuss how the results depend on the parameters mentioned above.

\subsection{Neutron emission during fusion-fission
process}


As discussed in reference \cite{ari04}, in order to simplify the
calculation, we assume again that the kinetic energy does not
dissipate during the approaching process, that is, $E_{diss}=0$
MeV. In subsection~3.2.2 below, the sensitivity of $E_{diss}$ is
discussed. In the three-dimensional Langevin calculation, we
prepare 1,000 trajectories, and at $t=0$, each trajectory starts
from the point of contact, which is defined as
$R_{touch}=R_{p}+R_{t}$, where $R_{p}$ and $R_{t}$ are the radii
of the projectile and the target, respectively.

The level density parameter $a_{n}$ for a certain nuclear shape
has been estimated by classical methods \cite{toke81}.
Fundamentally, we need $a_{n}$ for various deformations of the
nucleus. According to reference \cite{toke81}, the value of
$a_{n}$ increases with increasing deformation of the nucleus. In
order to simplify the calculation, we use two different values of
$a_{n}$, that is, the value inside the fusion box defined in the
deformation space as $\{z < 0.5, \delta < 0.3, |\alpha | < 0.3\}$,
and that outside the region, which are denoted by $a_{n}^{sph}$
and $a_{n}^{def}$, respectively. The excitation energy dependence
of $a_{n}$ is not considered. The neutron binding energy $B_{n}$
also depends on the deformation, but here we use the constant
value given for the ground state, because the fluctuation of $B_n$
in the entire deformation space relevant to the present
calculation is $\pm$0.3MeV. The influence of the fluctuation of
$B_{n}$ is discussed further in subsection~3.2.4. We chose the
values of $a_{n}^{sph}$ and $a_{n}^{def}$ so as to reproduce the
features of the experimental data, which are $a_{n}^{sph}=A/12.0$
and $a_{n}^{def}=A/10.0$.


\begin{figure}[h]
\centerline{
  \includegraphics[height=0.39\textheight]{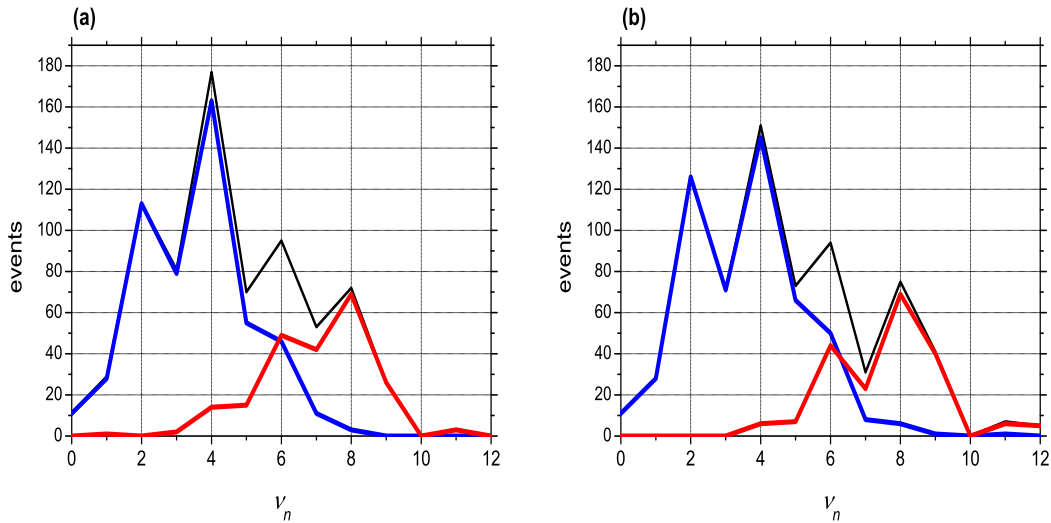}}
  \caption{Pre-scission neutron multiplicity given by our theoretical calculation
   in the reaction $^{58}$Ni+$^{208}$Pb at $E^{*}$=185.9 MeV. The
   neutron multiplicity from the QF process and the FF process are
   denoted by the blue and red line, respectively. The black line
   denotes the total processes.
  (a) in the case of $a_{n}^{sph}=A/12.0$ and $a_{n}^{def}=A/10.0$
  (b) in the case of $a_{n}^{sph}=A/14.0$ and $a_{n}^{def}=A/10.0$.}
\end{figure}


\begin{figure}
\centerline{
\includegraphics[height=.43\textheight]{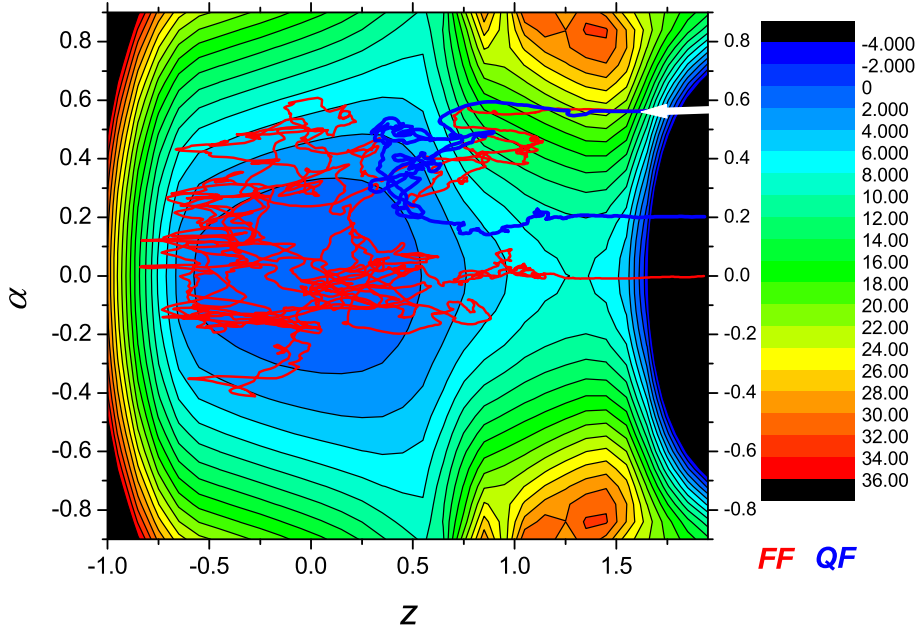}}
  \caption{Sample trajectories projected onto $z-\alpha$
$(\delta=0)$ plane at $E^{*}=185.9$ MeV in the reaction
$^{58}$Ni+$^{208}$Pb. The trajectories of the QF and the FF
processes are denoted by blue and red lines, respectively. The
potential energy surface is presented by the liquid drop model in
nuclear deformation space for $^{266}$Ds. The arrow denotes the
injection point of the reaction.}
\end{figure}

\begin{figure}
\centerline{
\includegraphics[height=.43\textheight]{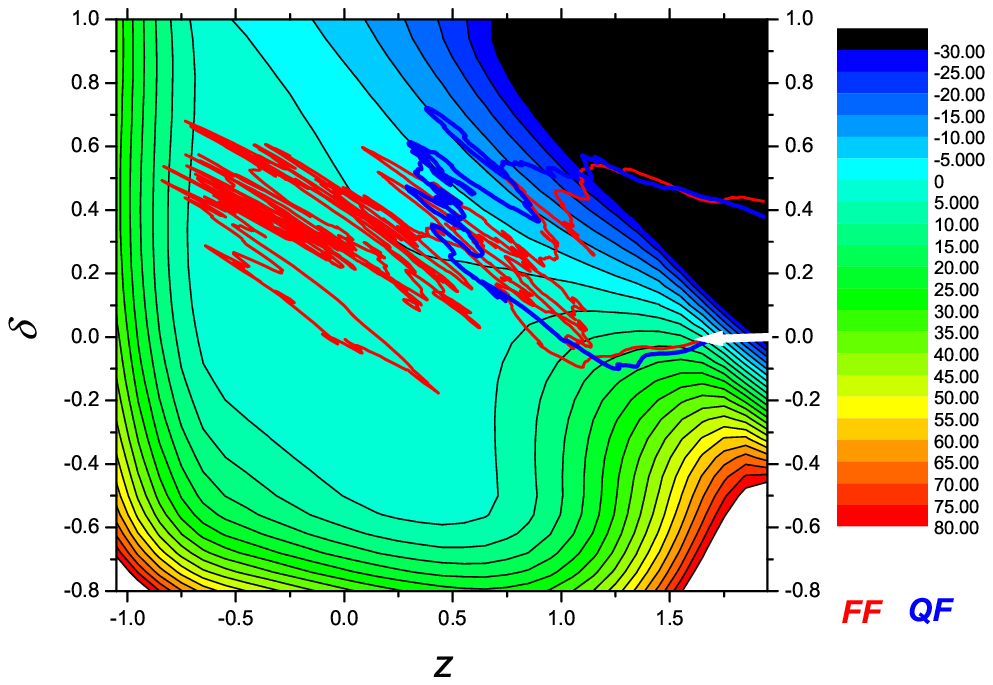}}
  \caption{The same samples of the trajectory as in Fig.~3, but which are
   projected onto $z-\delta$ $(\alpha=0)$ plane.}
\end{figure}

For the QF process, we plot the pre-scission neutron multiplicity
in correlation with fission fragments with mass numbers greater
than $\frac{A}{2}-30$ and less than $\frac{A}{2}+30$. To classify
the trajectory of the FF path, the definition of the fusion area
in the deformation space is very important. Here, we define the
fusion area (fusion box) as the inside of the fission saddle point
in the system described above. The idea behind the definition of
the fusion box is the same as that in reference \cite{ari04}. The
FF trajectory is identified as that which enters the fusion box.

The distribution of the pre-scission neutron multiplicity
$\nu_{n}$ calculated by our model is shown in Fig.~2(a) with the
level density parameters $a_{n}^{sph}=A/12.0$ and
$a_{n}^{def}=A/10.0$. The pre-scission neutron multiplicity for
the QF and the FF processes are denoted by the blue and the red
lines, respectively. The black line shows the total multiplicity
of each process.
As another example, we show in Fig.~2(b) the result using the
different level density parameter set, which are
$a_{n}^{sph}=A/14.0$ and $a_{n}^{def}=A/10.0$.

We can see clearly the two components which come from the QF
process and the FF process. It is shown that for the large neutron
multiplicity it originates from the FF path, and on the other hand
for small neutron multiplicity it comes from the QF path. This
means that the pre-scission neutron multiplicity has a strong
correlation with dynamical paths.


 On the pre-scission neutron multiplicity, the odd-even
 oscillations appear clearly, due to the odd-even effect on neutron
 binding energies. In the system, the initial number of neutrons is
 even, so that the probability of emitting an even number of
 neutrons is larger. The calculations show the similar structure
 observed in the experimental measurements in Fig.~1.


Next, we discuss the details on this calculation. Figures~3 and 4
show the potential energy surface of the liquid drop model for
$^{266}$Ds on the $z-\alpha$ $(\delta=0)$ plane and $z-\delta$
$(\alpha=0)$ plane, respectively, in the case of $l=0$. This
potential energy surface is calculated using the two-center shell
model code \cite{suek74,iwam76}. The contour lines of the
potential energy surface are drawn at steps of 2~MeV. Similar to
the deformation parameters of nuclear shape described in reference
\cite{ari04}, $z$ is defined as $z=z_{0}/(R_{CN}B)$, where
$R_{CN}$ denotes the radius of the spherical compound nucleus. The
parameter $B$ is defined as $B=(3+\delta)/(3-2\delta)$.

In Figs.~3 and 4, the position at $z=\alpha=\delta=0$ corresponds
to a spherical compound nucleus. The injection point of this
system is indicated by the arrow. The top of the arrow corresponds
to the point of contact in the system. We start the calculation of
the three-dimensional Langevin equation at the point of contact,
which is located at $z=1.575, \delta=0.0, \alpha=0.564$. All
trajectories start at this point with momentum in the initial
channel. That is to say, the initial velocity is directed in only
the $z$ direction, and the components of the initial velocities
along the $\delta$ and $\alpha$ directions are both assumed to be
zero. The sample trajectories of the QF process and the FF process
are shown in Figs.~3 and 4. The trajectories are projected onto
the $z-\alpha$ plane $(\delta=0)$ in Fig.~3 and $z-\delta$ plane
$(\alpha=0)$ in Fig.~4. The trajectory of the QF process and the
FF process are denoted by blue line and red line, respectively.

We define the travelling time $t_{trav}$ as a time duration during
which the trajectory moves from the point of contact to the
scission point. Figure 5 shows the distribution of travelling time
$t_{trav}$. The $t_{trav}$ from the QF and FF processes are
denoted by the solid and dashed lines, respectively. On the FF
process, as we can see in Figs.~3 and 4, the trajectory is trapped
in the pocket around the spherical region. The trajectory spends a
relatively long time in the pocket and it has a large chance to
emit neutrons. In average, the time duration spending in the
pocket is $7\times 10^{-20}$sec. On the other hand, the trajectory
of the QF process reaches to the scission point quickly. So, it
has not so much chance to emit neutrons. The mean travelling time
for the QF process is approximately $2\times 10^{-20}$sec.

\begin{figure}
\centerline{
  \includegraphics[height=.37\textheight]{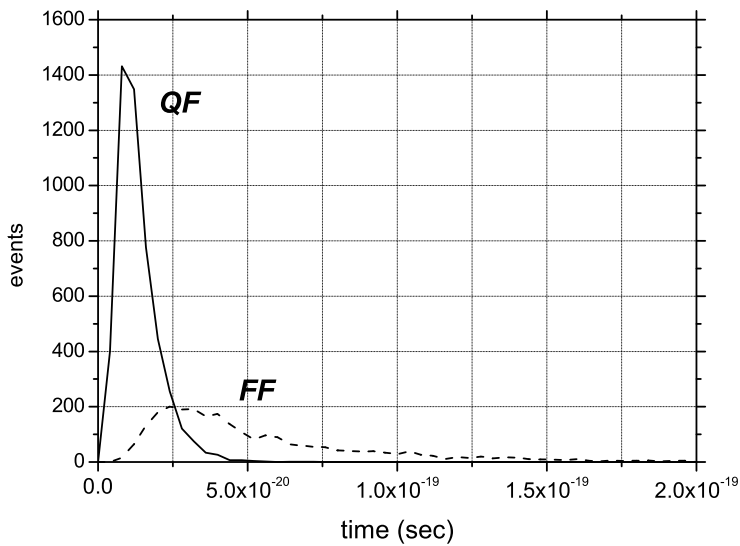}}
  \caption{The distribution of travelling time $t_{trav}$ in the
   reaction $^{58}$Ni+$^{208}$Pb at $E^{*}=185.9$ MeV. The $t_{trav}$ from the QF
   and FF processes are denoted by the solid and dashed lines, respectively.}
\end{figure}


Actually the distribution of $t_{trav}$ for each trajectory is
clearly distinguished by the two processes as shown in Fig.~5. The
time scale of the FF process is about 3 or 4 times longer than
that of the QF process. We can see in Figs.~3 and 4 that the
length of QF path is shorter than that of the FF path. On the
other hand, the FF trajectory is trapped at the potential pocket
near spherical region, so that the traveling time becomes longer.


\begin{figure}
\centerline{
  \includegraphics[height=.37\textheight]{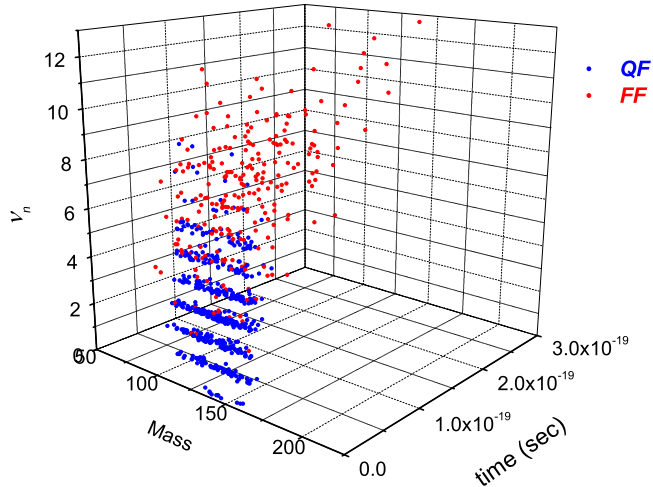}}
  \caption{The relation of the travelling time $t_{trav}$, the mass distribution of fission
  fragments and pre-scission neutron multiplicity $\nu_{n}$. The blue and red
dots are denoted the events from the QF and FF processes,
respectively. }
\end{figure}

It is interesting to investigate the correlation between the
travelling time and a number of events for neutron multiplicity.
Figure 6 shows the relation of the travelling time $t_{trav}$ and
pre-scission neutron multiplicity $\nu_{n}$. The blue dots and red
dots are denoted the events for the QF and the FF processes,
respectively. As shown in Fig.~6, the longer the travelling time
is, the larger the pre-scission neutron multiplicity becomes. Also
in the figure, the correlation between the mass distribution of
fission fragments and the pre-scission neutron multiplicity is
shown. The events from the QF process are distributing around
$\nu_{n}=2n, 3n$ and $4n$, and in the time axis they are
distributing around $t_{trav} \sim 1.5\times 10^{-20}$ sec. In the
case of the FF process, due to longer $t_{trav}$, the events are
scattered around $\nu_{n}=6n,7n$ and $8n$. Since the mass
distribution of fission fragments shows symmetric in this reaction
system due to high incident energy, we can not clearly see the
correlation between the mass fragment and the neutron
multiplicity.




\subsection{Parameter dependence of the model}

Our model has several parameters which are not confirmed
theoretically at present. We discuss the sensitivity of the
parameters of our model and show clearly the role of each
parameter. As the essential factors that influence the neutron
multiplicity, two categories are considered. They are concerned
with the travelling time of the trajectory and with the rate of
neutron emission. The former comprises the strength of the
friction force and the dissipation energy $E_{diss}$ at the point
of contact. The latter comprises the level density parameter
$a_{n}$ and the neutron binding energy $B_{n}$.

\subsubsection{Friction force $\gamma$}


\begin{figure}
\centerline{
  \includegraphics[height=.37\textheight]{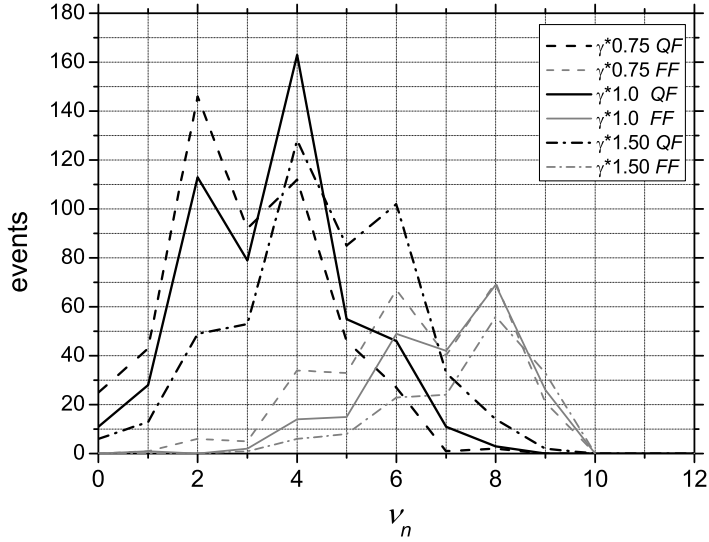}}
  \caption{The neutron multiplicity calculated using
one-body friction tensor multiplied by factors 0.75, 1.0 and 1.5,
as denoted by the dashed, solid and dashed-dotted lines,
respectively. The neutron multiplicity of the QF and FF processes
is dented by the black and gray lines, respectively.}
\end{figure}

In the dynamical calculation, the friction force plays a very
important role. The friction force influences mainly the
travelling time of the trajectory. Moreover, the rate of
dissipation at which the relative kinetic energy is converted into
the intrinsic energy and the strength of fluctuation in the
trajectory are controlled by the magnitude of the friction force.
In order to elucidate the sensitivity of the one-body friction
tensor to the trajectory calculation, the pre-scission neutron
multiplicity is shown in Fig.~7 for the modified strength of the
one-body friction tensor. The multiplied factors are 0.75, 1.0 and
1.5, and the corresponding results are shown by the dashed, solid
and dashed-dotted lines, respectively. The neutron multiplicities
from the QF and FF processes are dented by the black and gray
lines, respectively.

For large friction with the factor of 1.5, the trajectory speed is
slower than that in the case of friction with the factor of 1.0.
Due to longer $t_{trav}$, the neutron multiplicity $\nu_{n}$
becomes larger. As a result, the peaks of neutron multiplicity
from the QF process and the FF process shift to larger
multiplicity (to the right in Fig.~7). On the other hand, in the
case of the factor of 0.75 (smaller friction), the trajectory
speed is faster and the neutron multiplicity distribution shifts
to smaller $\nu_{n}$.

\begin{figure}
\centerline{
  \includegraphics[height=.37\textheight]{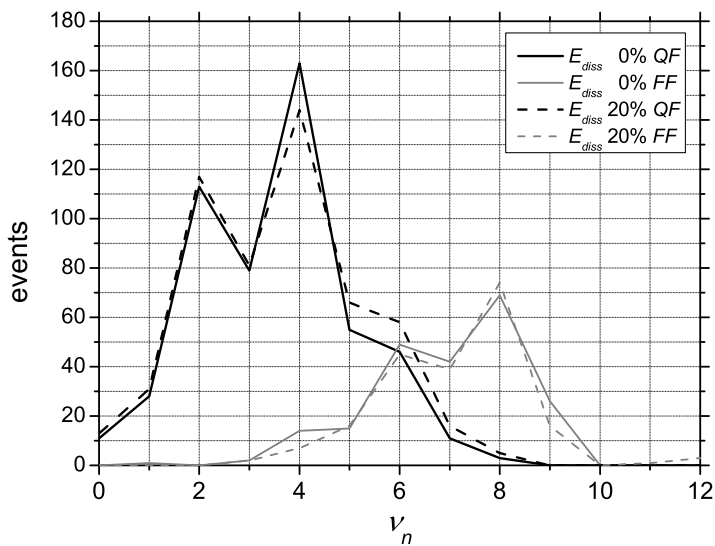}}
  \caption{The neutron multiplicity in the cases of $E_{diss}=0$ and 32.4 MeV,
  as denoted by solid and dashed lines, respectively. The black and gray
lines denote the neutron multiplicity from the QF and FF
processes, respectively.}
\end{figure}

\subsubsection{Dissipation energy $E_{diss}$}


As the initial condition of the Langevin calculation, the
intrinsic energy and relative kinetic energy of nuclei at the
point of contact are very important. However, in the
superheavy-mass region, it is an open question as to how much of
the relative kinetic energy of the colliding system dissipates
into intrinsic energy during the approaching process. Although we
assume no energy dissipation $E_{diss}$ at the point of contact in
our model calculation, in order to elucidate  the sensitivity of
$E_{diss}$, we deal with it at that point as a parameter. We have
discussed this subject on the basis of an analysis of the mass
distribution of fission fragments in reference \cite{ari03a}.
Here, we discuss how the energy loss $E_{diss}$ affects the
neutron multiplicity.

Since the initial momentum in the $-z$ direction strongly affects
the trajectory, $E_{diss}$ is expected to play a substantial role.
Figure~8 shows the neutron multiplicity in the cases of
$E_{diss}=0$ and 32.4 MeV (20\% of kinetic energy dissipates),
which are denoted by solid and dashed lines, respectively. The
black and gray lines denote the neutron multiplicity from the QF
and FF processes, respectively.


In the case of $E_{diss}$=32.4 MeV, due to the smaller initial
momentum in the $-z$ direction, the number of trajectories that
reach the spherical area decreases. However, due to the odd-even
effect on the neutron binding energies, such a variance of initial
momentum is too small to significantly change the neutron
multiplicity distribution.

Quantitatively, the mean value of $\nu_{n }$ for the QF process
increases from 3.53 to 3.60 when 20\% of the kinetic energy is
converted to the internal one during the approaching process. On
the contrary, in the same situation, the mean value of $\nu_{n }$
for the FF process decreases from 7.94 to 7.87.

\subsubsection{Level density parameter $a_{n}$}


\begin{figure}
\centerline{
  \includegraphics[height=.37\textheight]{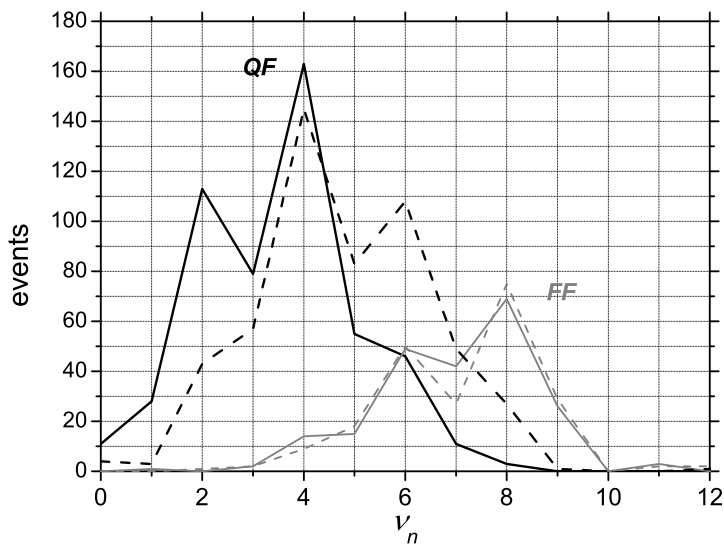}}
  \caption{The neutron multiplicity calculated using $a_{n}=A/12.0$ in
  the entire deformation space, denoted by the dashed lines. The
  solid lines denote the calculation using $a_{n}^{sph}=A/12.0$ and
   $a_{n}^{def}=A/10.0$. The black and gray lines denote the QF
   and FF processes, respectively.}
\end{figure}

In the statistical model calculation, the level density parameter
$a_{n}$ is very important. The lifetime of neutron emission is
strongly affected by $a_{n}$, because $a_{n}$ appears in the
exponent in the formula of the decay width, as shown by Eq.~(7).

First, we use the constant value $a_{n}$ in the entire deformation
space. The value of $a_{n}$ is chosen so as to reproduce the
experimental data of the neutron multiplicity from the spherical
region, that is, to reproduce the peak which is located at
$\nu_{n}$=8, where $a_{n}=A/12.0$ is used. The neutron
multiplicity in this case is shown by the dashed lines in Fig.~9.
The black and gray lines denote the neutron multiplicity from the
QF and FF processes, respectively. We can see that the value at
$\nu_{n}=6$ is rather large, which is not consistent with the
present experimental data.

According to reference \cite{toke81}, $a_{n}$ depends on the
deformation. As we did in the previous section, we use a different
value inside the fusion box and outside of it,
$a_{n}^{sph}=A/12.0$ and $a_{n}^{def}=A/10.0$, respectively. We
also plot this result in Fig.~9 as the solid lines. Due to the
larger value of $a_{n}^{def}$ in the deformation space compared
with $a_{n}^{sph}$, the emission rate of neutrons on the QF path
decreases.

\subsubsection{Neutron binding energy $B_{n}$}


\begin{figure}
\centerline{
  \includegraphics[height=.36\textheight]{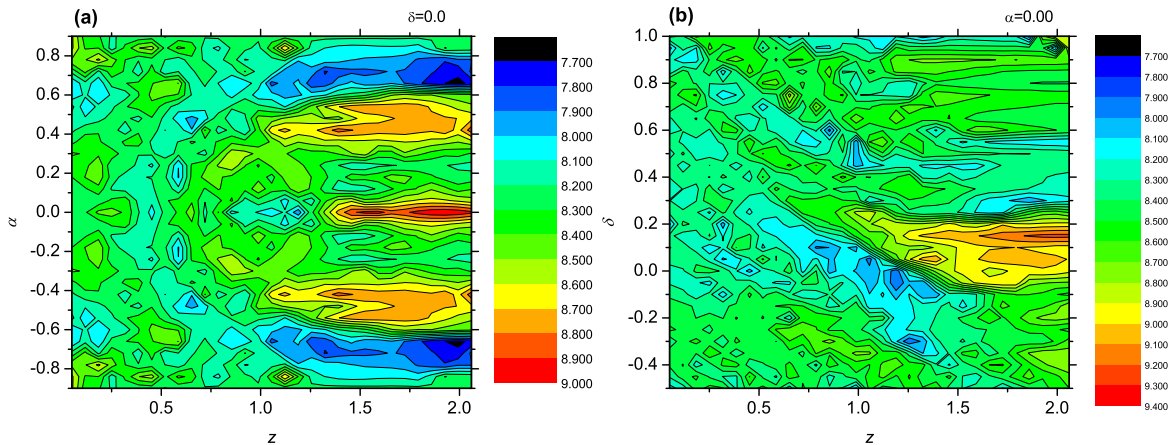}}
  \caption{The neutron binding energy $B_{n}$ of $^{266}$Ds on the deformation
  space. (a) $z-\alpha$ $(\delta=0)$ plane.
  (b) $z-\delta$ $(\alpha=0)$ plane.}
\end{figure}

Another parameter that strongly influences the decay width for
neutron emission is the neutron binding energy, $B_{n}$, of the
composite system. As shown in Eq.~(7), $B_{n}$ influences the
lifetime of neutron emission in the same manner as does $a_{n}$.
$B_{n}$ depends on the deformation of the nucleus. Figure~10(a)
and (b) shows $B_{n}$ of $^{266}$Ds on the $z-\alpha$ ($\delta=0$)
plane and $z-\delta$ ($\alpha=0$) plane, respectively. Here, we
use the two-center shell model code \cite{suek74,iwam76}. We can
see that in the whole deformation space, the neutron binding
energy fluctuates by about $\pm0.3$ MeV comparing with that for
spherical compound nucleus.

We investigate the $B_{n}$ dependence of the neutron multiplicity.
In the same way as in the case of $a_{n}$, we use a different
value of $B_{n}$ inside the fusion box ($B_{n}^{sph}$) and outside
of it ($B_{n}^{def}$). $B_{n}^{sph}$ means the neutron binding
energy of the compound nucleus. Figure~11 shows the neutron
multiplicity for three different sets of $B_{n}^{def}$, that is to
say, $B_{n}^{def}=B_{n}^{sph}$, $B_{n}^{def}=B_{n}^{sph}-0.3$ and
$B_{n}^{def}=B_{n}^{sph}+0.3$, as denoted by the solid, dashed and
dashed-dotted lines, respectively. The black and gray lines denote
the QF and FF processes, respectively.
The larger value of $B_{n}$ causes the emission rate of neutrons
to decrease. As a result, the distribution of the neutron
multiplicity shifts to smaller multiplicity (to the left in
Fig.~11). The large value of $B_{n}$ contributes the neutron
multiplicity in the same manner as the large value of $a_{n}$.

\begin{figure}
\centerline{
  \includegraphics[height=.37\textheight]{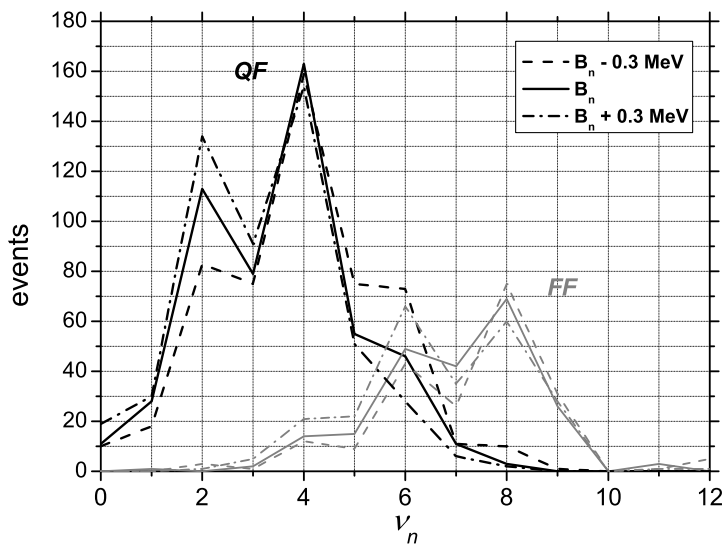}}
  \caption{The neutron multiplicity calculated using $B_{n}^{def}=B_{n}^{sph},
  B_{n}^{sph}-0.3,$ and $B_{n}^{sph}+0.3$, as denoted by the solid, dashed and
  dashed-dotted lines, respectively. $B_{n}^{sph}$ means the neutron binding
  energy of the compound nucleus.
  The black and gray lines denote the QF and FF processes,
   respectively. }
\end{figure}

\section{Summary}

We introduced the effect of pre-scission neutron emission into the
three-dimensional Langevin calculation, that is to say, we
combined the Langevin code with the statistical code. We applied
our model to the investigation of the whole fusion-fission
dynamical path in the reactions $^{58}$Ni+$^{208}$Pb at
$E^{*}$=185.9 MeV \cite{dona99}.

Our model is successfully applied to distinguish between the FF
process and the QF process in terms of the pre-scission neutron
multiplicity. We can see clearly the two components of the neutron
multiplicity, which come from the QF process and the FF process.
It is found that the peak for the large neutron multiplicity
originates from the FF path, and on the other hand the peak for
small neutron multiplicity comes from the QF path. The
pre-scission neutron multiplicity has a strong correlation with
dynamical path and is a powerful tool for investigating
fusion-fission mechanism.

We also discussed the sensitivity of the parameters of our model
and showed clearly the role of each parameter. As the essential
factors that influence the neutron multiplicity, we investigated
the friction force $\gamma$, dissipation energy $E_{diss}$, level
density parameter $a_{n}$ and neutron binding energy $B_{n}$.


This study will be the cornerstone for further study which
develops the methods of the classification between the FF process
and the DQF process. The D\'{e}MoN group already has measured the
pre-scission neutron multiplicity in reactions,
$^{48}$Ca+$^{208}$Pb, $^{48}$Ca+$^{244}$Pu and
$^{58}$Fe+$^{248}$Cm at $E^{*}\sim 40$ MeV \cite{mate04,nata04}.
In the next study, we would like to analyze these reactions.

The authors are grateful to Professor Yu.~Ts.~Oganessian,
Professor M.G.~Itkis, Professor V.I.~Zagrebaev for their helpful
suggestions and valuable discussion throughout the present work.
The authors thank Dr. S.~Yamaji and his collaborators, who
developed the calculation code for potential energy with a
two-center parameterization. This work has been in part supported
by INTAS projects 03-01-6417.



\bibliographystyle{aipproc}   

\bibliography{sample}

\IfFileExists{\jobname.bbl}{}
 {\typeout{}
  \typeout{******************************************}
  \typeout{** Please run "bibtex \jobname" to optain}
  \typeout{** the bibliography and then re-run LaTeX}
  \typeout{** twice to fix the references!}
  \typeout{******************************************}
  \typeout{}
 }

\end{document}